# Generalized virtual wave reconstruction for vibrothermography: Overcoming the wavefront-free behavior and quantification challenges in the diffusion-wave field

Pengfei Zhu[a,*], Julien Lecompagnon[a], Mathias Ziegler[a], Clemente Ibarra-Castanedo[b], Xavier Maldague[b]

[a] *Bundesanstalt für Materialforschung und -prüfung (BAM), 12200, Berlin, Germany*

[b] *Department of Electrical and Computer Engineering, Computer Vision and Systems Laboratory (CVSL), Laval University, Quebec G1V 0A6, Canada*

[*]Corresponding author: pengfei.zhu@bam.de

## Abstract

Wavefront-free behavior and the resulting quantification difficulties are intrinsic limitations of vibrothermography due to the diffusive nature of thermal fields. This work proposes a generalized virtual wave reconstruction framework to address the absence of propagation features in thermal diffusion-wave fields and its impact on quantitative defect characterization. Unlike conventional virtual wave formulations restricted to Dirac-type excitations, the proposed approach establishes a rigorous spatiotemporal mapping between diffusion-wave fields and virtual wave fields under arbitrary heat-generation conditions without simplifying assumptions on thermo-mechanical coupling. The resulting ill-posed inversion problem is solved using truncated singular value decomposition (T-SVD) and the alternating direction method of multipliers (ADMM) to enhance numerical stability and suppress noise amplification. Numerical simulations demonstrate that the reconstructed virtual wave fields recover propagation characteristics absent in temperature distributions, leading to improved defect boundary definition, contrast enhancement, and depth-resolved analysis. Experiments on CFRP laminates validate the robustness of the approach and show significantly improved signal-to-noise ratio, spatial clarity, and defect size estimation compared with conventional thermographic processing methods.

## 1. Introduction

Vibrothermography, also referred to as sonic infrared or ultrasound-stimulated thermography, is a non-destructive evaluation (NDE) technique that exploits the conversion of mechanical vibration energy into localized heat at defect sites [1]. In contrast to conventional ultrasonic inspection, which relies on the interpretation of wave reflections and scattering patterns [2], vibrothermography observes defect-induced thermal transients captured by an infrared camera, thereby reducing the need for extensive signal interpretation and enabling rapid, full-field visualization of crack activity [3]. However, there are several limitations which significantly impede the further application of vibrothermography techniques. For instance, the mechanical coupling [4] and excitation conditions [5], defect types [6], and surface conditions [7] significantly affect the detectability of defects in vibrothermography. In addition, due to the ambiguous and complex heating mechanism [8], it is extremely difficult to achieve quantitative evaluation of defects.

Common excitation devices in vibrothermography include piezoelectric [9] and electromagnetic [10] shakers, which provide controlled low-frequency or broadband mechanical vibration for stimulating crack face interactions. Piezo-ceramic actuators are frequently used for localized, high-efficiency energy input, enabling compact and easily integrated excitation setups. For ultrasonic stimulation, both air-coupled [11] and water-coupled [12] ultrasonic transducers offer flexible coupling configurations, with water coupling typically used when high-power or efficient energy transfer is needed. In addition, ultrasonic welders [13], owing to their ability to deliver high-intensity vibrations at well-defined frequencies, are widely utilized for rapid, high-energy excitation, especially in laboratory demonstrations of crack heating mechanisms. These diverse excitation systems allow vibrothermography to be tailored to different materials, geometries, and defect types, supporting both high-power and low-power inspection modalities.

The excitation methods in vibrothermography can be categorized as ultrasound lock-in thermography (ULT) [14], ultrasound burst phase thermography (UBP) [15], frequency-modulated ultrasonic excitation [16], and ultrasonic sweep thermography [17]. Ultrasound lock-in thermography (ULT) uses low-frequency amplitude-modulated ultrasound to stimulate nonlinear defect behavior such as friction or crack face rubbing. The periodic modulation of the ultrasonic amplitude causes periodic heat generation at the defect, turning it into a local thermal source. An infrared camera records the surface temperature over multiple modulation cycles, and lock-in processing, via Fourier analysis or sinusoidal fitting, extracts the temperature amplitude and phase at each pixel. The resulting amplitude and phase images suppress emissivity variations and highlight zones of energy dissipation, allowing defects to be visualized with high contrast. Ultrasound burst phase thermography (UBP) builds on the principles of ULT but replaces sinusoidal excitation with short, high-energy ultrasonic bursts. Instead of tracking a periodic thermal wave, the infrared camera records the heating and subsequent cooling transient produced by each burst. Phase information is then extracted from the local spectral content of the temperature-time signal, providing defect contrast similar to ULT but with greater robustness to coupling variations and significantly shorter measurement times. Because the defect-related signal is concentrated within a limited frequency band, UBP further reduces broadband noise and improves signal-to-noise ratio. Burst excitation also produces a broad spectral response, enabling depth estimation through Fourier analysis at different frequencies. With its high sensitivity and fast acquisition, UBP is particularly effective for detecting delaminations in carbon fiber reinforced polymers (CFRP) and other layered composites. Frequency-modulated ultrasonic excitation is used to overcome problems associated with monofrequency stimulation, where the excitation frequency may coincide with a structural resonance and produce standing waves. Such standing wave patterns can generate misleading thermal signatures due to hysteretic losses at vibration maxima, while actual defects may lie at nodal positions and therefore remain undetected. By driving the specimen with two or more frequencies simultaneously, or by continuously modulating the frequency of a sinusoidal signal, the standing-wave field is superimposed with propagating waves. This reduces blind spots at nodes and improves sensitivity across the entire inspection area. Ultrasonic sweep thermography applies a controlled frequency sweep, typically over a narrow ultrasonic band such as 20-25 kHz, to examine how a component responds thermally at different excitation frequencies. During the sweep, the vibration amplitude is kept constant, and the slowly varying excitation can last from a few seconds to several minutes. This approach is particularly effective for crack detection in metals, where defect activation can depend strongly on the excitation frequency. The frequency at which the largest lock-in temperature amplitude occurs can be identified from the sweep, enabling the selection of an optimal excitation frequency. When combined with burst-type acquisition, the resulting infrared image sequence can be Fourier-transformed at multiple thermal wave frequencies, producing phase images that offer depth-resolved information from a single measurement.

The heating mechanism in vibrothermography is more complex than other infrared thermography techniques [18,19,20,21]. Researchers performed numerous experiments for different defects to find the fundamental reason about the heat generation [22]. Although this is still an unsolved mystery now [3], researchers have already understood the main reason. Heat generation in vibrothermography arises from the material's mechanical response under cyclic ultrasonic or low-frequency vibration [23]. In general, the thermal response of a solid can be separated into reversible thermoelastic heating and irreversible dissipative heating [24]. Because vibrothermography typically operates at stress levels below the plastic threshold, plastic strain effects are negligible, and the dominant heat mechanisms include thermoelastic effect, internal friction and hysteresis losses, and crack-interface friction and clapping. Thermoelastic effect causes reversible temperature fluctuations due to elastic compression and expansion. It contributes to small periodic thermal signal. Under cyclic loading, real materials deviate slightly from perfectly elastic behavior. The hysteresis loop between loading and unloading represents energy lost per cycle, which is converted into heat. This dissipation depends on material properties, vibration frequency, and internal damping. Defects generate strong localized heating because the elastic properties inside a crack differ rom those of the intact material. When crack faces rub, slide, or repeatedly close and open, frictional losses and micro-impact (clapping) events produce concentrated heat at the defect interface. In conclude, the heating mechanism in vibrothermography is a multi-physical field coupling problem. The heating behavior also located at different temporal scales. In addition, it is extremely difficult to determine all material's (or damage) parameters. Therefore, directly solving the analytical solution considering all heating phenomenon is hard. Researchers tried to neglect the microscopic heating effect and consider only a heat source at the internal defect location. In this case, the simplified model can be solved [25]. However, this is inaccurate and cannot provide effective information of defect size or depth. Researchers also tried to use experimental methods such as laser doppler vibrometry (LDV) to study the relationship between the excitation frequency and heating efficiency [26]. The quantitative analysis is still a hard issue until now.

To achieve higher signal-to-noise ratio, i.e., improving the defect contrast, researchers mainly transferred the knowledge from general optical excitation thermography [27]. For instance, the most commonly used pulsed phase thermography [28] and principal component thermography [29] were employed in results from vibrothermography. Recently, with the development of deep learning, researchers also tried to apply unsupervised autoencoder to reduce the infrared image noise [30]. The unsupervised networks not only provide a novel strategy for denoising but also eliminating the data hungry problem in non-destructive testing fields. In addition, to improve the spatial resolution, researchers also apply the transfer learning-based generative adversarial network (GAN) to infrared images [31]. However, there is few open literatures discussing the specific image/signal processing techniques on vibrothermography.

In this work, a generalized virtual wave method is proposed for the image processing and quantitative detection in vibrothermography. This generalized virtual wave is built on the conventional virtual wave concept and extends it from the specific Dirac pulse excitation to arbitrary excitation without any simplification during the theoretical modeling. Two mathematical algorithms, including truncated singular-value decomposition (T-SVD) and alternating direction method of multi-pliers (ADMM), were employed to numerically solve the ill-posed inversion problem for transforming the thermal diffusion-wave field to elastic (virtual) wave field. As a result, the inherently wavefront-free behavior and quantification challenges are solved. The numerical simulation between thermal diffusion-wave field, elastic wave field, and reconstructed virtual wave field were conducted for comparison. Finally, the experiments were performed to validate the robustness and feasibility of the proposed virtual wave method.

## 2. Theory

### 2.1. Heat generation in vibrothermography

The complete temperature field $T(x,t)$ obeys the standard heat conduction equation with a general time-dependent heat source $Q(t)$:

$$\rho C \frac{\partial T}{\partial t} - \nabla(k \nabla T) = Q(t) \tag{1}$$

where $\rho$ is the density, $C$ is the specific heat, $k$ is the thermal conductivity, and $Q(t)$ is the total mechanical-to-thermal energy dissipation caused by ultrasonic excitation. The total heat generation is the sum of two physically distinct mechanisms:

$$Q(t) = Q_{\text{int}}(t) + Q_{\text{friction}}(t) \tag{2}$$

where the internal heating $Q_{\text{int}}(t)$ can be given as

$$Q_{\text{int}}(t) = \frac{1}{2}\omega \text{Re}[\varepsilon(x,t) \cdot D\dot{\varepsilon}(x,t)] \tag{3}$$

where $\varepsilon(x,t)$ is the strain field, $\dot{\varepsilon}(x,t)$ is the strain-rate field, and $D$ is the elasticity matrix. The internal heating $Q_{\text{int}}(t)$ originates from the area of the stress-strain hysteresis loop under cyclic micro-vibration at the ultrasonic frequency. For cracks or delaminations, the most important heating mechanism heating mechanism is contact friction, caused by crack-face sliding, rubbing, and clapping. The second term of Eq. (2) can be given as:

$$Q_{\text{friction}}(t) = \int_\Gamma [\mu_d + (\mu_s - \mu_d)e^{-\eta|v(x,t)|}]p_c(x,t)v(x,t)dA \tag{4}$$

where $\mu_s$ and $\mu_d$ are static and dynamic friction coefficients, $p_c(x,t)$ is the contact pressure, $v(x,t)$ is the relative sliding velocity between crack surfaces, and $\eta$ is the transition parameter from static to dynamic friction. The crack-face frictional heating represents the instantaneous mechanical energy converted to heat on the crack surfaces. Combining all mechanisms, the full transient model is:

$$\rho C \frac{\partial T}{\partial t} - \nabla(k \nabla T) = \frac{1}{2}\omega \text{Re}[\varepsilon(x,t) \cdot D\dot{\varepsilon}(x,t)] + \int_\Gamma [\mu_d + (\mu_s - \mu_d)e^{-\eta|v(x,t)|}]p_c(x,t)v(x,t)dA \tag{5}$$

### 2.2. Conventional virtual wave concept

Conventional virtual wave concept is based on the Dirac pulse excitation [32]. The thermal diffusion equation can be given as:

$$\left(\nabla^2 - \frac{1}{\alpha}\frac{\partial}{\partial t}\right)T(\boldsymbol{r},t) = -\frac{1}{\alpha}T_0(\boldsymbol{r})\delta(t) \tag{6}$$

where $\alpha = k/(\rho C)$ is the thermal diffusivity. The external heat incident is assumed as the initial condition instead of boundary condition. The equivalence between these two boundary-value problems was validated by Zhu et al [33]. However, in real case, external heat source is impossible to be a Dirac pulse. In general, the heating period is at the level of millisecond. Therefore, Eq. (6) is suitable to describe an ideally point heating source. The wave equation describes the acoustic pressure $p(\boldsymbol{r},t)$ as a function of space $\boldsymbol{r}$ and time $t$ and can be written as:

$$\left(\nabla^2 - \frac{1}{c^2}\frac{\partial^2}{\partial t^2}\right)p(r,t) = -\frac{1}{c^2}\frac{\partial}{\partial t}p_0(r)\delta(t) \tag{7}$$

where $c$ is the speed of sound, $p_0(r)$ is the initial pressure distribution just after the short excitation pulse. A virtual wave $T_{\text{virt}}$ is defined as the same wave equation with the initial temperature distribution $T_0(r)$ and an arbitrary chosen $c$:

$$\left(\nabla^2 - \frac{1}{c^2}\frac{\partial^2}{\partial t^2}\right)T_{\text{virt}}(r,t) = -\frac{1}{c^2}\frac{\partial}{\partial t}T_0(r)\delta(t) \tag{8}$$

Both Eq. (6) and Eq. (8) can be transformed into the frequency domain:

$$(\nabla^2 - \sigma(\omega)^2)\theta(r,\omega) = -\frac{1}{\alpha}T_0(r) \tag{9}$$
$$(\nabla^2 - \psi(\omega)^2)\theta_{\text{virt}}(r,\omega) = -\frac{i\omega}{c^2}T_0(r) \tag{10}$$

where $\sigma(\omega)^2 \equiv i\omega/\alpha$ and $\psi(\omega) = \omega/c$ are wavenumber for thermal diffusion-wave field and virtual wave field. Replacing $\omega$ by $-ic\sigma(\omega)$ in Eq. (10), the relationship between thermal diffusion-wave field and virtual wave field in the frequency domain reads:

$$\theta(r,\omega) = \frac{c}{\alpha\sigma(\omega)}\theta_{\text{virt}}(r,-ic\sigma(\omega)) \tag{11}$$

Therefore, the temperature in time-domain can be given as:

$$T(r,t) = \frac{1}{2\pi}\int_{-\infty}^{\infty} T_{\text{virt}}(r,t')K(t,t')dt' \tag{12}$$

where $K(t,t') \equiv \frac{c}{\sqrt{\pi\alpha t}}\exp\left(-\frac{c^2 t'^2}{4\alpha t}\right)$ for $t > 0$. Eq. (12) constructs the relationship between thermal diffusion-wave field and virtual wave field in the time domain. This is also a Fredholm integral equation of the first kind. By solving Eq. (12), the virtual wave field can be obtained.

*2.3. Generalized virtual wave reconstruction theory*

As mentioned in the Section 2.2, the conventional virtual wave method is limited to Dirac-type pulsed excitations. When the heating source deviates from an ideal Dirac pulse, Eq. (12) fails to accurately reconstruct the corresponding virtual wave field. The mechanism of vibrothermography is inherently complex, as it involves a coupled structure-thermal system: the ultrasonic excitation operates at kilohertz frequencies, whereas heat conduction occurs on the order of hertz. Existing analytical treatments often simplify vibrothermography by modeling it purely as a heat conduction process with an internal heat source. Such simplifications introduce significant discrepancies from the real scenario, because they neglect the underlying heat generation mechanism induced by the mechanical excitation.

In this work, we do not attempt to impose any a prior assumptions or simplifications on the heat generation model in Eq. (5). Instead, the heat source is treated in its most general form, defined at arbitrary spatial location $r$ and time $t$. The governing thermal diffusion equation is thus expressed as:

$$\left(\nabla^2 - \frac{1}{\alpha}\frac{\partial}{\partial t}\right)T(r,t) = -\frac{1}{k}Q(r,t) \tag{13}$$

Analogously, a virtual wave field $T_{\text{virt}}(r,t)$ can be introduced, defined as the solution of the corresponding wave equation:

$$\left(\nabla^2 - \frac{1}{c^2}\frac{\partial^2}{\partial t^2}\right)T_{\text{virt}}(r,t) = -\frac{1}{c^2}Q(r,t) \tag{14}$$

Both Eq. (13) and Eq. (14) can be transformed into the frequency domain, yielding:

$$(\nabla^2 - \sigma(\omega)^2)\theta(r,\omega) = -\frac{1}{k}\tilde{Q}(r,\omega) \tag{15}$$
$$(\nabla^2 - \psi(\omega)^2)\theta_{\text{virt}}(r,\omega) = -\frac{1}{c^2}\tilde{Q}(r,\omega) \tag{16}$$

Replacing $\omega$ in Eq. (16) by $-ic\sigma(\omega)$ establishes a direct relation between the thermal and virtual wave fields:

$$\theta(\bm{r},\omega) = \frac{c^2}{k}\theta_{\text{virt}}(\bm{r},-ic\sigma(\omega))) \tag{17}$$

Therefore, the temperature field in the time domain can be reconstructed via a convolution integral:

$$T(\bm{r},t) = \int_{-\infty}^{\infty} T_{\text{virt}}(\bm{r},t')K(t,t')dt' \tag{18}$$

where the kernel $K(t,t') = \frac{c^2}{k}\frac{t'}{2\sqrt{\pi}(\alpha t)^{3/2}}\exp\left(-\frac{c^2 t'^2}{4\alpha t}\right)H(t)H(t')$, and $H(\cdot)$ denotes the Heaviside step function. Eq. (18) constitutes the generalized virtual wave reconstruction theory, which rigorously connects the thermal diffusion-wave field with the corresponding virtual wave field. Of note, this formulation is universally applicable to arbitrary heat sources, as no simplifying assumptions regarding the heat generation mechanism were made during its derivation. This provides a robust framework for reconstructing thermal fields in complex vibrothermography scenarios, bridging the gap between diffusive thermal transport and wave-based analysis.

## 3. Numerical reconstruction algorithms

Eq. (18) establishes a direct relationship between the temperature signal and the corresponding virtual wave signal at the same spatial location $\bm{r}$, which can be expressed mathematically as a Fredholm integral equation of the first kind. Since temperature measurements are inherently discrete in both time and space, Eq. (18) can be approximated in a discretized form:

$$\bm{T} = \bm{K}\bm{T}_{\text{virt}} \tag{19}$$

where $\bm{T}$ and $\bm{T}_{\text{virt}}$ are matrices representing the measured temperature signals and the virtual wave signals at discrete time steps, respectively. Direct inversion of the matrix $\bm{K}$ is ill-conditioned due to several factors: 1) Temporal spreading of thermal diffusion smooths the temperature evolution over time, analogous to how wave scattering broadens acoustic signals. This results in the columns of $\bm{K}$ being nearly linearly dependent. 2) Exponential attenuation of short-time features in the thermal response suppresses high-frequency components of the virtual wave, similar to the loss of high-frequency ultrasound signals in tissue. 3) Noise amplification in direct inversion would introduce unphysical oscillations in the reconstructed virtual wave signal, as small measurement errors are magnified by the ill-conditioned matrix. To address these challenges, we propose two numerical strategies for stable reconstruction of the virtual wave field, which mitigate the effects of rank deficiency and measurement noise.

### 3.1. Alternating direction method of multi-pliers (ADMM)

An iterative non-linear regularization technique, the alternating direction method of multi-pliers (ADMM), can be used to inversely reconstruct the virtual wave field $\bm{T}_{\text{virt}}$. The key advantages of ADMM lie in its ability to incorporate sparsity prior into the regularization process, which effectively suppresses artificial oscillations in the solution. In the context of vibrothermography, the thermal wave signal can reasonably be assumed to be sparse due to the limited number of material interfaces present in the sample. Under this assumption, the reconstruction problem can be formulated as the following minimization:

$$\min_{\bm{T}_{\text{virt}}} \frac{1}{2}\|\bm{K}\bm{T}_{\text{virt}} - \bm{T}\|_2^2 + \lambda_{\text{ADMM}}\|\bm{T}_{\text{virt}}\|_1 \tag{20}$$

Here, the first term enforces fidelity to the measured temperature data $\bm{T}$, while the second term imposes sparsity through the $\ell_1$-norm of $\bm{T}_{\text{virt}}$. The $\ell_1$-norm allows the solution to retain steep gradients, enabling sharper transitions in the virtual wave signal and producing a solution that is less oversmoothed compared to traditional Tikhonov regularization. The regularization parameter $\lambda_{\text{ADMM}}$ is determined using the L-curve criterion, which balances the trade-off between data fidelity and sparsity enforcement. For practical implementation, the discretized form of the reconstruction problem for two spatial dimensions $x$ and $y$ is given as in Eq. (18), where $\bm{T} \in \mathbb{R}^{N_t \times N_x \times N_y}$ is the observed temperature data, $\bm{K} \in \mathbb{R}^{N_t \times N_{t'}}$ is the discrete kernel, $\bm{T}_{\text{virt}} \in \mathbb{R}^{N_{t'} \times N_x \times N_y}$ is the reconstructed virtual wave field. The entries of the discrete kernel $\bm{K} = [K_{i,j}] \in \mathbb{R}^{N_t \times N_{t'}}$ can be calculated as:

$$K(i,j) = \frac{\tilde{c}^2}{\kappa}\frac{(j-1)}{2\sqrt{\pi}(\tilde{\alpha}i)^{3/2}}\exp\left(-\frac{\tilde{c}^2(j-1)^2}{4\tilde{\alpha}i}\right) \tag{21}$$

where the dimensionless speed of sound $\tilde{c} = c\frac{\Delta t}{\Delta z}$ and the dimensionless thermal diffusivity $\tilde{\alpha} = \alpha\frac{\Delta t'}{\Delta z^2}$ are defined in terms of the increments of the measurement time $\Delta t$, the virtual time $\Delta t'$, and the spatial depth $\Delta z$. This formulation allows for a robust numerical reconstruction of the virtual wave field from discrete temperature measurements while controlling noise amplification and preserving sharp features associated with material interfaces.

*3.2. Truncated singular-value decomposition (T-SVD)*

The discrete inverse problem can also be addressed using singular-value decomposition (SVD) of the kernel matrix $\boldsymbol{K}$:

$$\boldsymbol{K} = \boldsymbol{U\Sigma V}^T \tag{22}$$

where $U = (u_1, \ldots, u_N)$ and $V = (v_1, \ldots, v_N)$ are orthonormal matrices, and $\boldsymbol{\Sigma} = \text{diag}(\sigma_1, \ldots, \sigma_N)$ contains the singular values ordered as $\sigma_1 \geq \sigma_2 \geq \cdots \geq \sigma_N \geq 0$. Formally, the inverse problem can be expressed as:

$$\boldsymbol{T}_{\text{virt}} = \sum_{n=1}^{N} \frac{u_n^T \boldsymbol{T}}{\sigma_n} v_n \tag{23}$$

Due to the diffusive nature of thermal transport, high temporal-frequency components are exponentially suppressed, resulting in singular values that decay approximately as the inversion and amplify measurement noise, leading to instability. To stabilize the reconstruction, a truncation index $r$ is introduced, and the truncated inverse is defined as:

$$\boldsymbol{K}_r^{-1} = \sum_{n=1}^{r} \frac{1}{\sigma_n} v_n u_n^T \tag{24}$$

The corresponding reconstructed virtual wave field becomes:

$$\boldsymbol{T}_{\text{virt}}^{(r)} = \sum_{n=1}^{r} \frac{u_n^T \boldsymbol{T}}{\sigma_n} v_n \tag{25}$$

This truncation effectively removes components associated with $\sigma_n \ll \sigma_r$, which are dominated by noise. Physically, the diffusion kernel acts as a temporal Gaussian filter with width $\delta t \sim \frac{4\alpha t}{c^2}$, which limits the recoverable temporal bandwidth of the virtual-wave components to $\omega_{\max} \sim \frac{c^2}{4\alpha t}$. Therefore, the truncation index $r$ corresponds to a diffusion-limited temporal resolution, analogous to the diffraction limit in conventional wave imaging. Equivalently, the Picard condition implies $|u_n^T \boldsymbol{T}| \ll \sigma_n$ for large $n$. Stable inversion requires truncation before the measurement noise intersects the singular spectrum. From Eq. (18), short-time virtual-wave components are attenuated as $\exp\left(-\frac{c^2 t'^2}{4\alpha t}\right)$. The smallest resolvable virtual time satisfies $t'_{\min} \sim \sqrt{\frac{4\alpha t}{c^2} \ln \frac{1}{\text{SNR}}}$. Hence, truncated SVD (T-SVD) does not simply serve as a numerical regularization technique. It also enforces the thermodynamic irreversibility constraint inherent to diffusion. In matrix notation, the T-SVD inversion operator is:

$$\boldsymbol{T}_{\text{virt}}^{(r)} = \boldsymbol{V}_r \boldsymbol{\Sigma}_r^{-1} \boldsymbol{U}_r^T \boldsymbol{T} \tag{26}$$

where the subscript $r$ denotes restriction to the dominant singular subspace.

## 4. Simulation analysis

In this work, we do not study the efficiency of heat generation caused by cracks/delaminations shape or ultrasound excitation frequency. Instead, we focus on the transformation between thermal diffusion-wave field and virtual wave field. Therefore, the heat generation was simplified as the boundary heat source at elliptical delamination defect with semiaxis of 5 and 0.2 mm. The sample size is 40 mm × 5 mm. The material type is a stainless steel with the thermal conductivity of 16.2 W/(m·K), the density of 7900 kg/m³, and the heat capacity of 477 J/(kg·K). Except for the delamination boundary, all boundaries were set as the heat convection with a convection coefficient of 10 W/(m²·K). For the amplitude modulation vibrothermography, the internal heat source was set to $8 \times 10^3 \times [1 + \sin(2\pi f t)]$ where $f = 0.5$ Hz. The temperature response directly above the defect is shown in Fig. 1(a). It increases with a periodic wave variation. The internal temperature field evolution is shown in Fig. 1(b). The temperature field exhibits a periodic variation with an increasement of the DC component.

To compare with results in thermal diffusion-wave field, we also performed the simulation of elastic wave field. The solid mechanics module in Comsol was used. The structural load comes from the boundary load at the boundaries of delaminations. It was set to $1 \times 10^{-6} \times [1 + \sin(2\pi f t)]$ where $f = 0.5$ Hz. The displacement response directly above the defect is shown in Fig. 1(c). It becomes the overlapping of different sine waves caused by the reflection, scattering, and transmission from boundaries. The evolution of the displacement field is shown in Fig. 1(d). It is clear to find the wavefront and the propagation path of the elastic wave. However, the displacement field does not show a periodic distribution as the media (space) here is not periodic. The overlapping effect causes the complex spatial pattern of the displacement field.

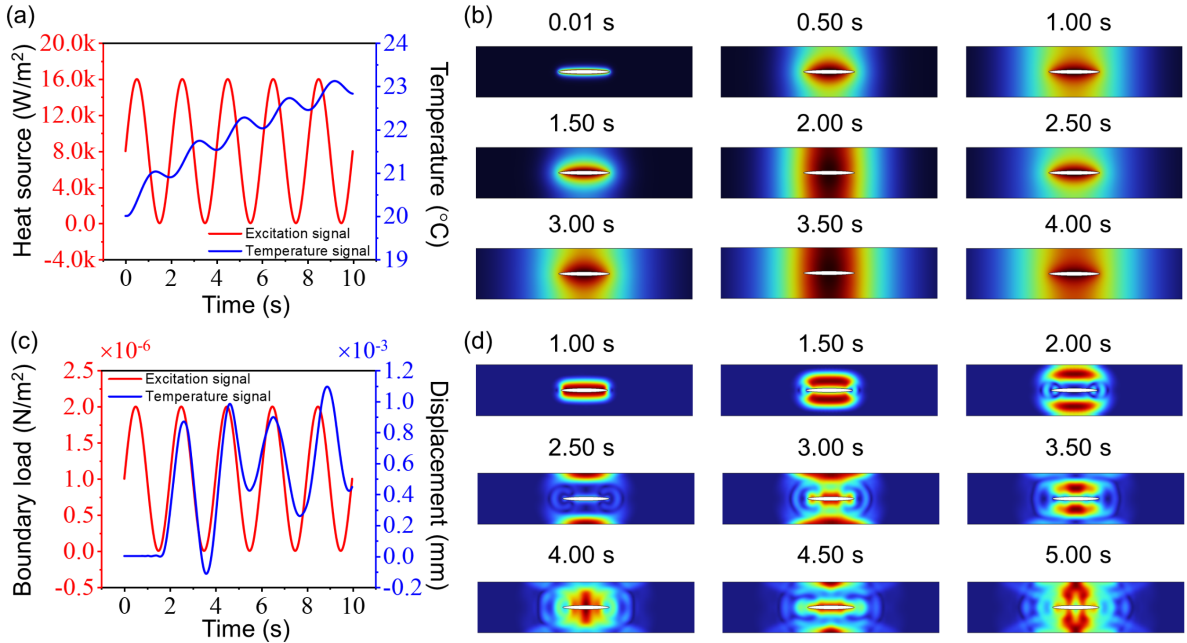

**Fig. 1.** Thermal diffusion-wave and elastic wave fields propagation: (a) Heat source and temperature response, (b) The evolution of internal temperature field, (c) Boundary load and displacement response, (d) The evolution of internal displacement field.

Due to the fact that we can only obtain the surface signals, the reconstruction of internal information is an ill-posed inverse problem. Therefore, many researchers proposed various inverse solving algorithms in both ultrasound [34] and infrared thermography fields [35]. To improve the defect contrast, signal modulation is a commonly used method. Here, the single-frequency sine signal was employed as the excitation. It is clear that the difference between temperature distribution and displacement distribution at surface is the boundary blurring, as shown in Fig. 2(a) and (b). This is caused by the wavefront-free nature of thermal diffusion-wave fields. Virtual wave reconstruction was employed to process the diffusion-wave field in order to restore the wavefront and achieve quantitative analysis. The reconstructed virtual wave field using T-SVD method is shown in Fig. 2(c). The primary feature between virtual wave field and thermal diffusion-wave field is that the signal is no longer accumulation due to the fact that the physical heat diffusion is a continuous Markov process. For lock-in (single-frequency) excitation, it is effective to analyze results in the frequency domain. The spectrum results of temperature, displacement, and virtual wave fields are shown in Fig. 2(d), (e), and (f). Due to the lateral diffusion, there no exactly sudden change at the boundary of delaminations in the frequency-domain temperature field. The artificial definition of defect boundary should be introduced in order to obtain the lateral delamination size. However, it will also introduce measurement uncertainty and errors since different experimental setup and tested samples will significantly affect the final results. This is also a core problem in the literatures of infrared thermography fields. Researchers only validated the effectiveness of their proposed methods on the dataset they present instead of proposing a general method [36,37,38]. In addition, in the delamination area, there are also noisy fringes in the frequency-domain thermal diffusion-wave field. According to results from frequency-domain displacement field, there are many interference fringes causing by the reflection and scattering of the boundary. Although displacement field results provide clear sudden change, they are significantly deviated the real boundary size of delaminations. In contrast, virtual wave field not only present high contrast and clear delamination boundaries, but also consistent with the real delamination size, as shown in Fig. 2(f).

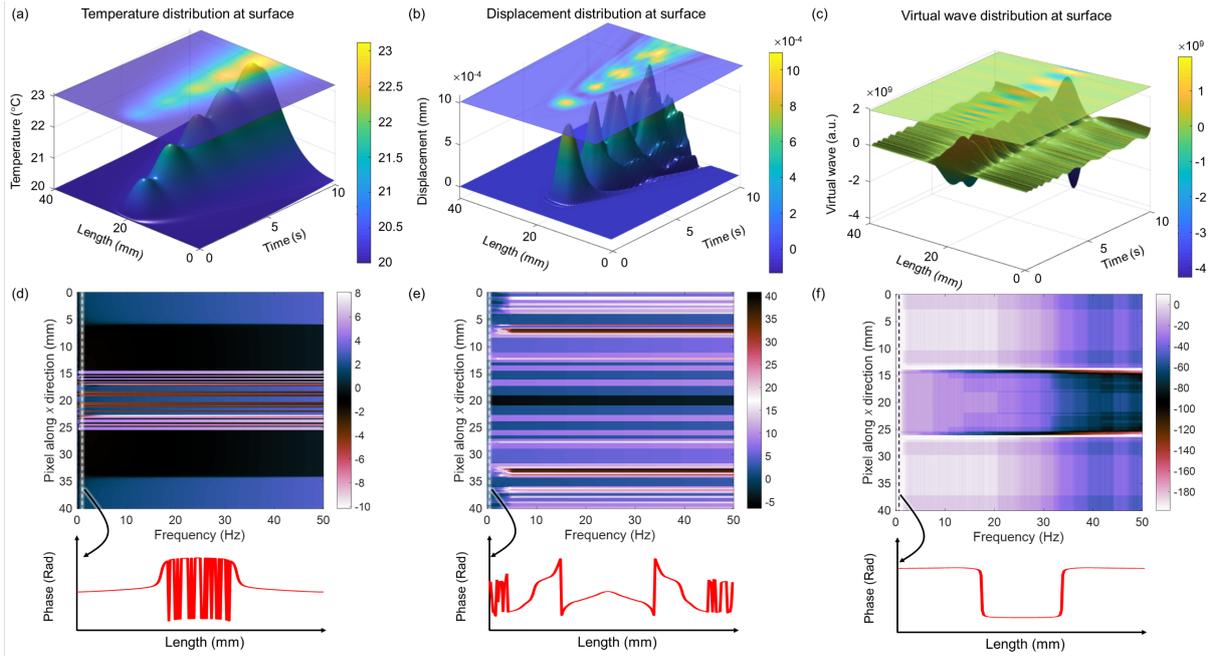

**Fig. 2.** Virtual wave reconstruction from thermal diffusion-wave to elastic wave fields propagation using T-SVD method: (a) Temperature distribution at surface, (b) Displacement distribution at surface, (c) Virtual wave distribution at surface, (d) Temperature field in the frequency domain, (e) Displacement field in the frequency domain, (f) Virtual wave field in the frequency domain.

To present the quantitative capability of the virtual wave reconstruction method, the similar simulations were performed for the heat transfer and elastic wave models. The same tested sample has three delamination defects located at different layers, where the center depths of these three delaminations are 3 mm, 2 mm, and 1 mm, respectively. The evolution of temperature field is shown in Fig. 3(a), while the evolution of elastic wave field is shown in Fig. 3(b).

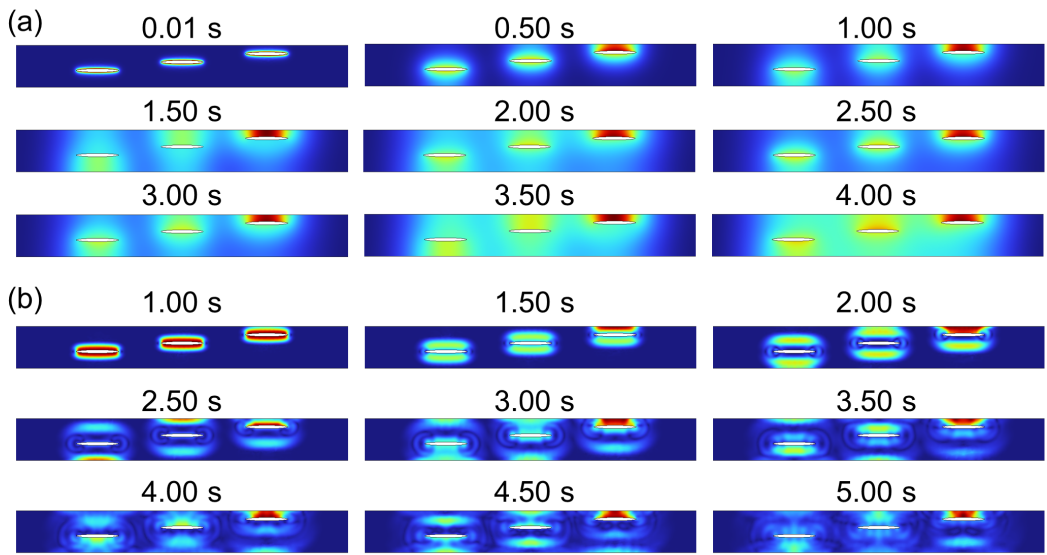

**Fig. 3.** Thermal diffusion-wave and elastic wave fields propagation: (a) The evolution of internal temperature field, (b) The evolution of internal displacement field.

The temperature distribution and displacement distribution at surface are shown in Fig. 4(a) and (b). Similarly, due to the gradient-driven, thermal diffusion-wave field exhibits serious blurring, while the displacement field presents clear wave-front. In particular, it is possible to find the difference of reaching time of internal wave from displacement field, while it is only possible to find the amplitude variation of different delamination areas. The reconstructed virtual wave field using ADMM method is shown in Fig. 4(c). After virtual wave inversion, it is clear to find the wavefront in virtual wave field. To further quantitatively detect the delamination depth, we selected the center points from three delamination areas, as shown in Fig. 4(d). For the elastic wave field, although we can observe the time delay of the internal wave, the depth corresponded by the peak time has large deviation from the real value. This is the reason why researchers always use the pulse excitation in ultrasonic testing for quantification. In the virtual wave field, the detected results are accurate except for the 3 mm delamination. This is because the effective information is significantly lost during the heat propagation from a deep area. Although ADMM algorithm is quite effective, it cannot perfectly restore the real information of the elastic wave field.

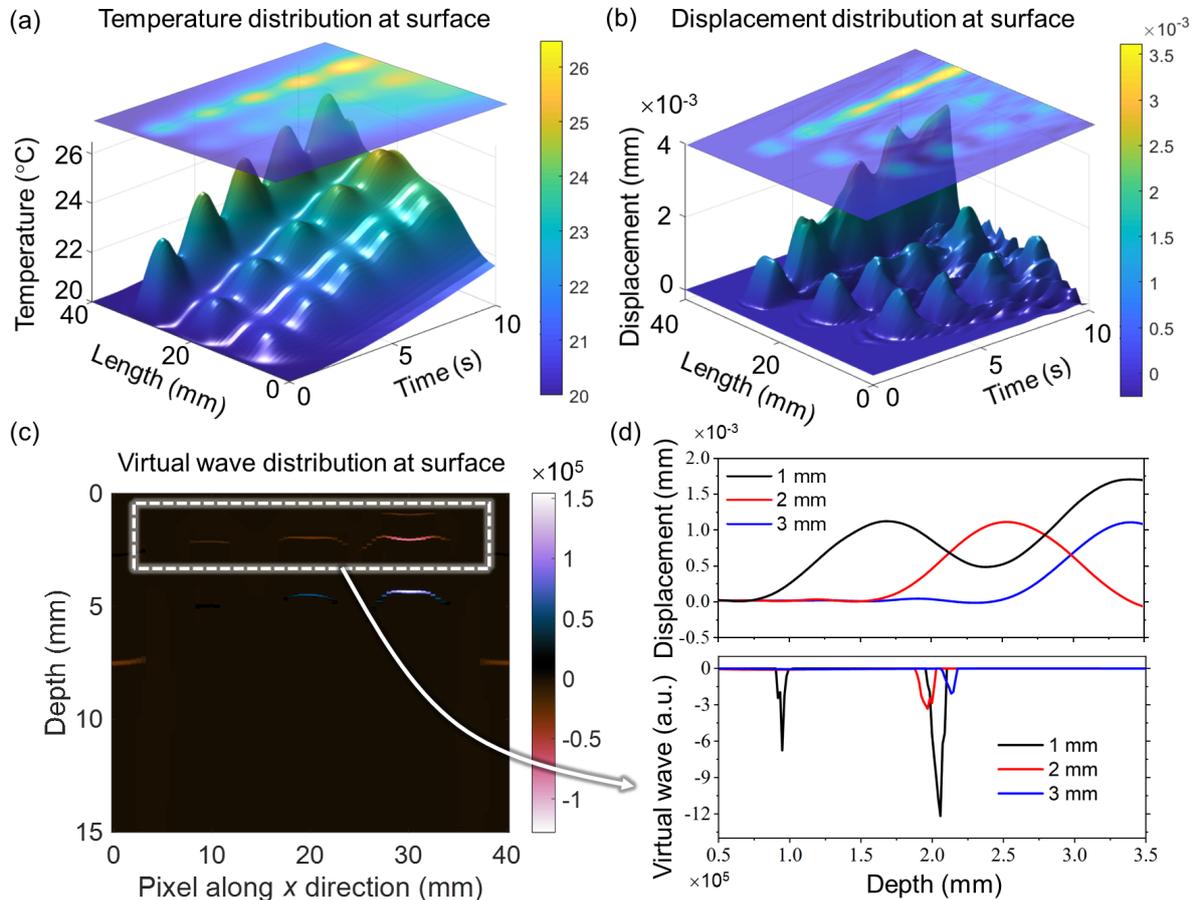

**Fig. 4.** Virtual wave reconstruction from thermal diffusion-wave to elastic wave fields propagation using ADMM method: (a) Temperature distribution at surface, (b) Displacement distribution at surface, (c) Virtual wave distribution at surface, (d) Depth inverse solving based on elastic wave field and virtual wave field.

## 5. Experimental setup

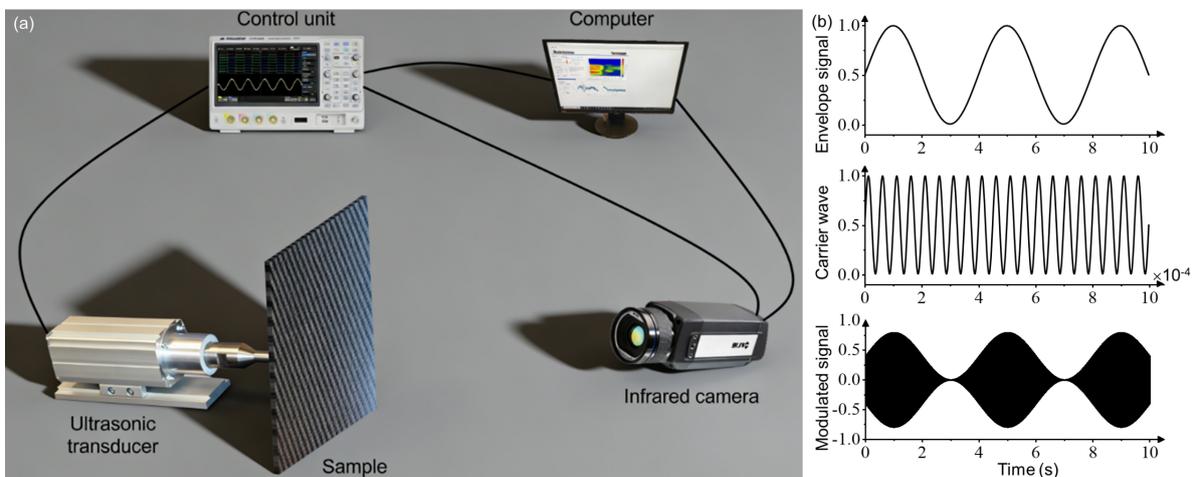

**Fig. 5.** The experimental setup for vibrothermography: (a) Schematic image, (b) Modulated signals.

A broadband ultrasound transducer (Branson 200b, 15-25 kHz) was employed to stimulate the specimen from the back side, while the thermal response was recorded from the front surface [39]. In this configuration, a lock-in vibrothermography approach was employed in a static setup, where internally generated heat is induced by mechanical energy input from an amplitude-modulated ultrasonic source, as shown in Fig. 5(b). The overall experimental setup is shown in Fig. 5(a). During testing, the sample was excited with a 25 kHz ultrasonic carrier wave modulated at frequencies of 0.125, 0.25, and 0.5 Hz. The resulting surface temperature field was captured using a high-speed mid-wave infrared (MWIR) camera (Telops FAST-IR 1000 MW, InSb, 3-5 μm, 320 × 256 pixels) operating at an acquisition frame rate of 100 Hz.

The tested sample is a ten-ply carbon fiber reinforced polymer (CFRP) laminate [40] containing 25 embedded square defects created using Teflon inserts of identical thickness but varying lateral dimensions and burial depth, as shown in Fig. 6. The thermal properties of the CFRP laminate are: thermal conductivity 0.8 W/(m·K), heat capacity 1200 J/(kg·K), and density 1600 kg/m$^3$. For Teflon, the thermal conductivity, heat capacity, and density are 0.25 W/(m·K), 1050 J/(kg·K), and 2170 kg/m$^3$, respectively. The laminate measures 300 mm × 300 mm × 2 mm. The 25 embedded inserts span a range of length-to-depth ratios between 1.8 and 75, enabling systematic evaluation of defect detectability.

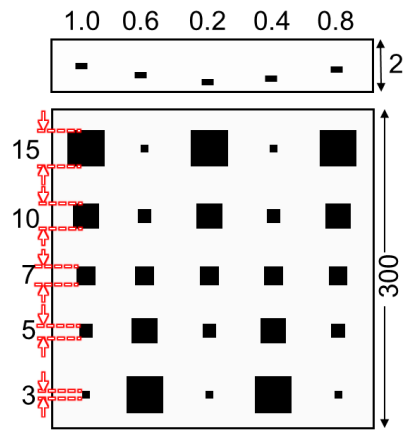

**Fig. 6.** The schematic image of the CFRP sample. Unit: mm.

## 6. Results and discussion

*6.1. Image processing based on conventional algorithms*

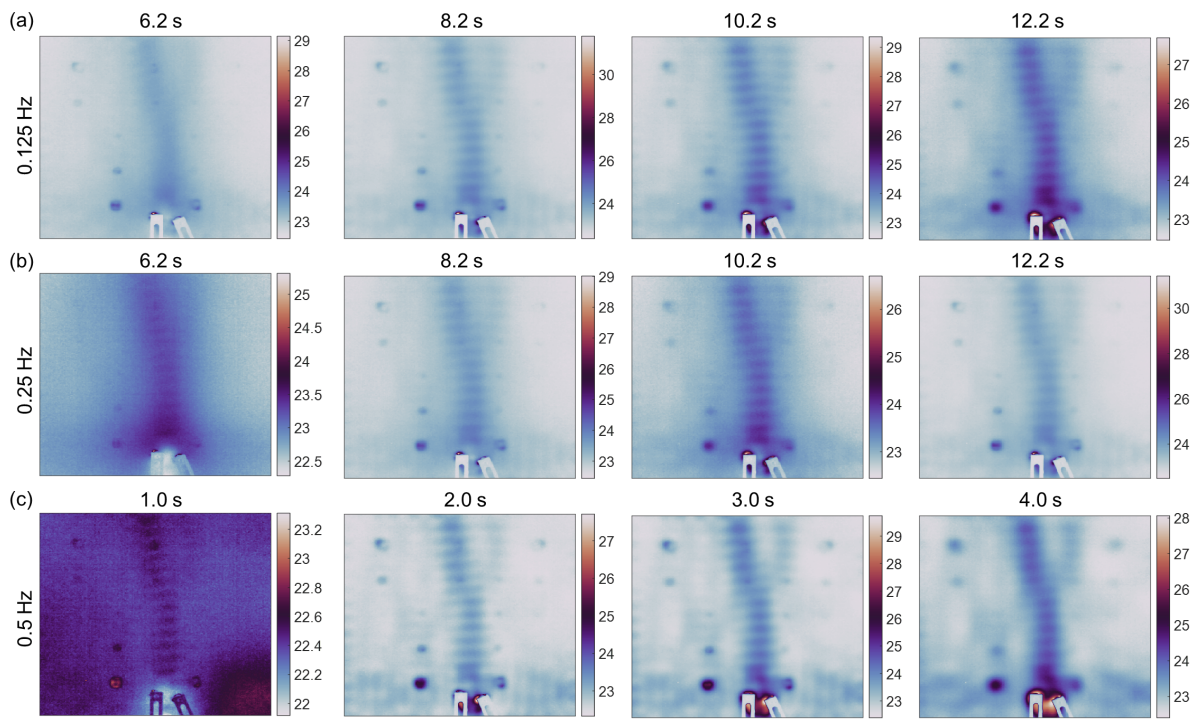

**Fig. 7.** The raw images under different envelope frequencies: (a) 0.125 Hz, (b) 0.25 Hz, (c) 0.5 Hz.

The raw images are shown in Fig. 7. Almost all defects with sizes below 7 mm cannot be detected in three envelope frequencies. It is clear to find that all detected results contain a "vertical heat band" and ripple patterns. This narrow vertical band of elevated temperature extends upward from the ultrasound coupling area. This feature is characteristic of vibrothermography in anisotropic composites such as CFRP. It is mainly caused by three physical mechanisms, i.e., directional propagation of ultrasonic energy in anisotropic CFRP [41], concentrated energy channel generated by non-uniform ultrasonic coupling [42], and path preferentially occurred by the layer-to-layer micro-slip and interface friction [43]. As we all know, CFRP has high attenuation through thickness but much lower attenuation and guided propagation in the plane of the laminate. As a results, when the ultrasonic transducer excites the back surface, the mechanical energy preferentially propagates along the laminate plan, which often aligned with the major fiber direction. The transducer contact area often introduces localized stress concentration, a directional vibration pattern (depending on horn geometry), and non-uniform coupling pressure. These effects cause mechanical energy to be injected into the laminate in a non-axisymmetric, directional manner. In addition, since the CFRP laminate has a layered structure, these frictional energy losses occur most strongly along a continuous path between plies, reinforcing the vertical heating band.

The periodic, wave-like thermal patterns are a direct thermal manifestation of the ultrasonic vibration field. In this work, the size of the CFRP laminate is 300 mm × 300 mm × 2 mm, which readily supports Lamb waves, flexural resonances, and in-plane standing waves. At 25 kHz, the wavelength of these guided modes falls within a few millimeters to a few centimeters. This produces periodic regions of maximum vibration amplitude (antinodes) and minimum amplitude (nodes). Since vibrothermography captures energy dissipation ($Q \propto \varepsilon^2$, where $\varepsilon$ is the strain amplitude), nodes and antinodes translate directly into alternating high- and low-temperature stripes. In addition, since the heating is modulated at 0.125-0.5 Hz, the captured signal represents only the portion of the thermal field synchronized with the modulation. Therefore, the spatial vibration pattern serves as the carrier, then lock-in extracts the periodic part of the temperature modulation. Finally, these cause clean, periodic ripple patterns.

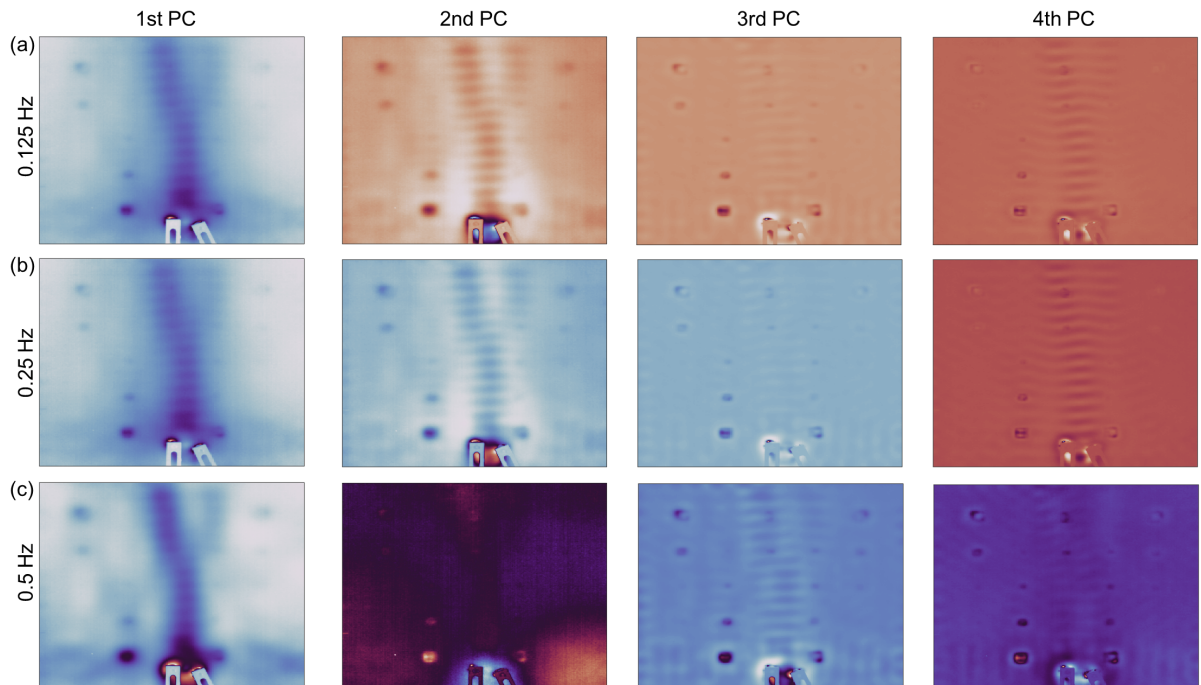

**Fig. 8.** The images processed by the PCT method under different envelope frequencies: (a) 0.125 Hz, (b) 0.25 Hz, (c) 0.5 Hz.

Only depending on the raw images cannot extract effective information since most of defects are hidden by the background noise. Therefore, it is necessary to introduce image processing methods. The inherent nature of thermal diffusion wave determines that the perfect inversion of internal structures of the tested sample is impossible. This is because the heat transfer obeys a parabolic partial differential equation. This equation smooths temperature fields over time. High-frequency spatial details decay exponentially, where small defects correspond to high-frequency components. It is difficult to directly transfer the knowledge of signal processing from ultrasonic or computer vision fields to infrared thermography field [44,45]. In this work, we employed three classic but quite effective image processing algorithms in infrared thermography field, including principal component thermography (PCT) [46], pulsed phase thermography (PPT) [47], and correlation method [48]. PCT is based on principal component analysis (PCA) to extract dominant thermal patterns from an infrared image sequence. The thermal image sequence is first reshaped into a 2D matrix $\mathbf{T} \in \mathbb{R}^{N \times M}$, where $N$ is the number of pixels and $M$ is the number of time frame. Singular value decomposition (SVD) is then applied,

$$\mathbf{T} = \mathbf{U\Sigma V}^T \tag{27}$$

where $\mathbf{U}$ and $\mathbf{V}$ are orthogonal matrices and $\mathbf{\Sigma}$ contains singular values. The first few principal components (columns of $\mathbf{U}$) are selected to reconstruct enhanced images, highlighting defect regions while suppressing noise. PPT analyses the thermal response in the frequency domain. For each pixel, the temporal temperature signal $T(t)$ is transformed via discrete Fourier transform (DFT),

$$\Theta(f) = \tan^{-1}\frac{\text{Im}[F(f)]}{\text{Re}[F(f)]} \tag{28}$$

$$F(f) = \sum_{t=0}^{M-1} T(t)e^{-i2\pi ft/M} \tag{29}$$

where $\Theta(f)$ is the phase at frequency $f$, and $M$ is the number of frames. Defects delay thermal wave propagation, producing phase contrast that is robust against surface emissivity variations and non-uniform heating. The correlation method evaluates the similarity between each pixel's temporal signal $T_i(t)$ and a reference signal $R(t)$ (e.g., theoretical or baseline response). The normalized correlation coefficient is calculated as

$$\rho_i = \frac{\sum_{t=0}^{M-1}(T_i(t)-\bar{T}_i)(R(t)-\bar{R})}{\sqrt{\sum_{t=0}^{M-1}(T_i(t)-\bar{T}_i)^2 \sum_{t=0}^{M-1}(R(t)-\bar{R})^2}} \tag{30}$$

where $\bar{T}_i$ and $\bar{R}$ are the mean values of the signals. Pixels with lower correlation indicate potential defects, providing a quantitative measure for defect detection.

The results processed by PCT method are shown in Fig. 8. The first four principal components (PCs) were extracted for the defect detection. Unfortunately, PCT method does not exhibit better defect contrast or explore new defects comparing with raw images in Fig. 7.

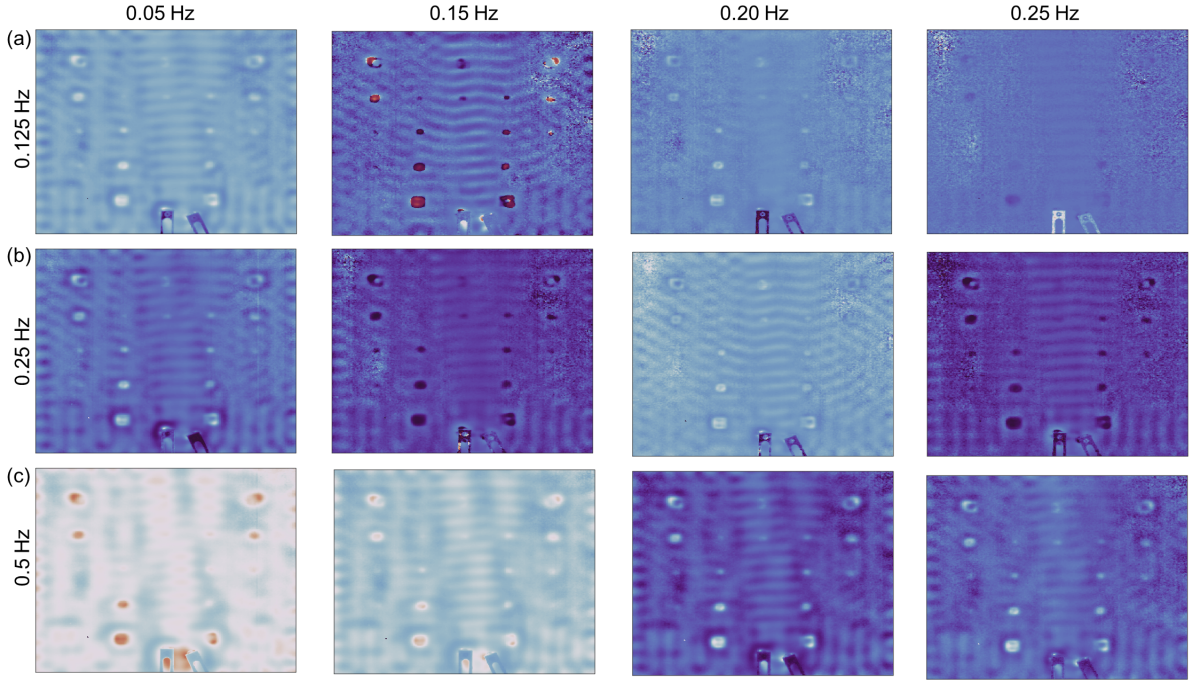

**Fig. 9.** The images processed by the PPT method under different envelope frequencies: (a) 0.125 Hz, (b) 0.25 Hz, (c) 0.5 Hz.

The results processed by PPT method are shown in Fig. 9. The four effective frequency components were extracted for the defect detection. It is clear to find that PPT method significantly improves the defect contrast comparing with results in Fig. 7 and Fig. 8. The vertical heat band is removed effectively by the phase thermogram. This is because the vertical heating strip is not modulated at the lock-in frequency and dominated by low spatial frequencies, resulting in almost no contribution in phase domain. In addition, originally hidden defects appear such as the 5 mm and 7 mm defects located at the left column, the 5 mm defect located at the second column, the 10 mm and 15 mm defects located at the central column, and the 7 mm defect located at the right column. Comparing these results in Fig. 9, the envelope frequency of 0.125 Hz exhibits the best detection capability, while the envelope frequency of 0.5 Hz shows the worst detection capability. Of note, the ripple-like patterns are enhanced by the phase thermogram. This is because these patterns originate from high-spatial-frequency ultrasonic modal vibrations whose envelope is modulated exactly at the reference frequency. These features produce strong, phase-coherent thermal oscillations, making them highly visible in the phase domain.

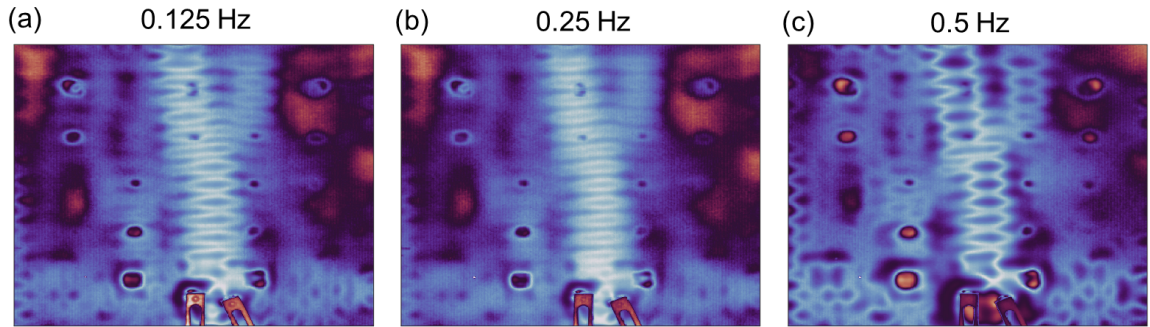

**Fig. 10.** The images processed by the correlation method under different envelope frequencies: (a) 0.125 Hz, (b) 0.25 Hz, (c) 0.5 Hz.

The results processed by correlation method are shown in Fig. 10. The correlation algorithm enhances thermal signals that are temporally synchronized with the ultrasonic modulation. As a result, the vertically elongated heating region-representing guided mechanical energy propagation and interlaminar friction-reappears with strong contrast, in contrast to PPT phase thermograms where this low-spatial-frequency component is largely suppressed. Meanwhile, the ripple-like patterns associated with standing ultrasonic modes become even more pronounced, as correlation selectively amplifies high-spatial-frequency, phase-coherent oscillations. Local defects appear as dark regions due to phase delay or attenuation in the modulated thermal response. Increasing modulation frequency (see Fig. 10(c)) sharpens these features due to reduced thermal diffusion length, yielding clearer modal fringes and more localized defect signatures. However, comparing with PPT and PCT methods, correlation algorithm demonstrates a worse detection capability since both vertical heating band and ripple-like patterns are not of interest in this work.

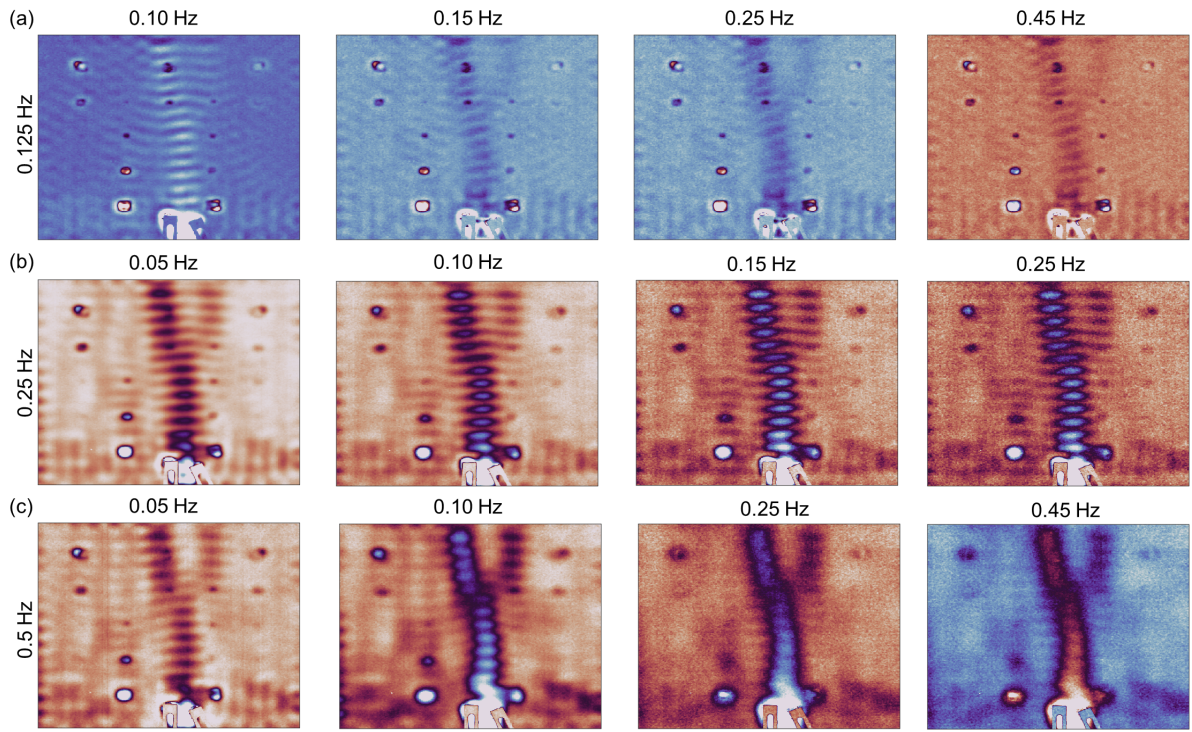

**Fig. 11.** The images processed by the virtual wave method under different envelope frequencies: (a) 0.125 Hz, (b) 0.25 Hz, (c) 0.5 Hz.

*6.2. Image processing based on virtual wave method*

The results processed by virtual wave method are shown in Fig. 11. The T-SVD was employed to solve the ill-posed inversion problem. Then, the phase information was extracted from the virtual wave results. The thermal diffusion-wave field was converted into approximate wave-like field. As a consequence, the periodic ripple-like patterns generated by ultrasonic modal vibrations become significantly more pronounced than in raw images. The vertical heating channel remains visible but becomes sharply localized, reflecting the removal of diffusion blur. Defects exhibit strong, well-defined thermal signatures due to the combination of gradient amplification and coherent demodulation. The virtual wave method effectively suppresses diffusive components while amplifying coherent, wave-like thermal oscillations, yielding superior contrast for both modal patterns and subsurface defects.

**Table 1**
Number of detected defects using different image processing techniques.

| Frequency | Raw image | PCT | PPT | Correlation | Virtual wave |
|---|---|---|---|---|---|
| 0.125 Hz | 6 | 6 | **15** | 10 | 12 |
| 0.25 Hz | 6 | 6 | **15** | 10 | 12 |
| 0.5 Hz | 10 | 10 | **13** | 10 | 10 |

To quantitative compare different image processing methods, several indices were used. The first used index is the number of detected defects, as shown in Table 1. In the raw images, only six defects can be detected under 0.125 Hz and 0.25 Hz excitations, while PPT method reveals 15 defects. Under 0.5 Hz excitation, except for the PPT method, all image processing methods do not reveal hidden defect on the basis of raw images. According to the total number of detected defects, we can conclude that they follow a rank of PPT (43) > Virtual wave (34) > Correlation (30) > PCT (22) = Raw image (22). Although this rank slightly depends on the subjective judgment, both PPT and Virtual wave methods are superior to other methods apparently.

The second used index is the signal-to-noise ratio (SNR), which is a commonly used index in NDE fields. The mathematical expression of SNR can be written as [49]:

$$\text{SNR} = 20 \log_{10} \frac{\mu_d - \mu_s}{\sigma_s} \quad (31)$$

where $\mu_d$ and $\mu_s$ are the mean values of defect and sound areas, respectively. $\sigma_s$ is the standard deviation of the sound area.

To accurately compare different methods, the 15 mm defect at the second column was selected as the defect area, and the surround area was selected as the background area. Of note, the defect area and background area are fixed for all compared results. The quantitative evaluation for different image processing techniques based on SNR is shown in Table 2. It is clear that correlation method has the lowest SNR as it enhances the vertical heating channel and ripple-like patterns in Fig. 10. Although PPT can reveal new hidden defects, its results have relatively worse SNR than raw images. According to the average value of different excitation results, we can conclude that they follow a rank of Virtual wave (28.28 dB) > PCT (27.97 dB) > Raw image (24.28 dB) > PPT (19.82 dB) > Correlation (10.78 dB).

**Table 2**
Quantitative evaluation for different image processing techniques based on SNR.

| Frequency | Raw image | PCT | PPT | Correlation | Virtual wave |
|---|---|---|---|---|---|
| 0.125 Hz | 25.97 dB | 29.17 dB | 24.80 dB | 12.19 dB | **33.43 dB** |
| 0.25 Hz | 24.82 dB | **32.26 dB** | 16.73 dB | 7.37 dB | 28.03 dB |
| 0.5 Hz | 22.07 dB | 22.49 dB | 17.92 dB | 12.77 dB | **23.38 dB** |

The third index is the measurement of defect size. Similarly, the 15 mm defect at the second column was selected for comparison. We simultaneously measured the lateral and vertical sizes of this defect. The full width of the half maximum (FWHM) was selected as the final measured value. The quantitative evaluation results of the defect size are shown in Fig. 12. Of note, the FWHM varies with the time in raw images [50]. We selected the raw image at the peak time for measuring the defect size. It is obvious to find that the temperature profile is unstable in both the correlation and the PPT methods. Therefore, they have relatively low credibility.

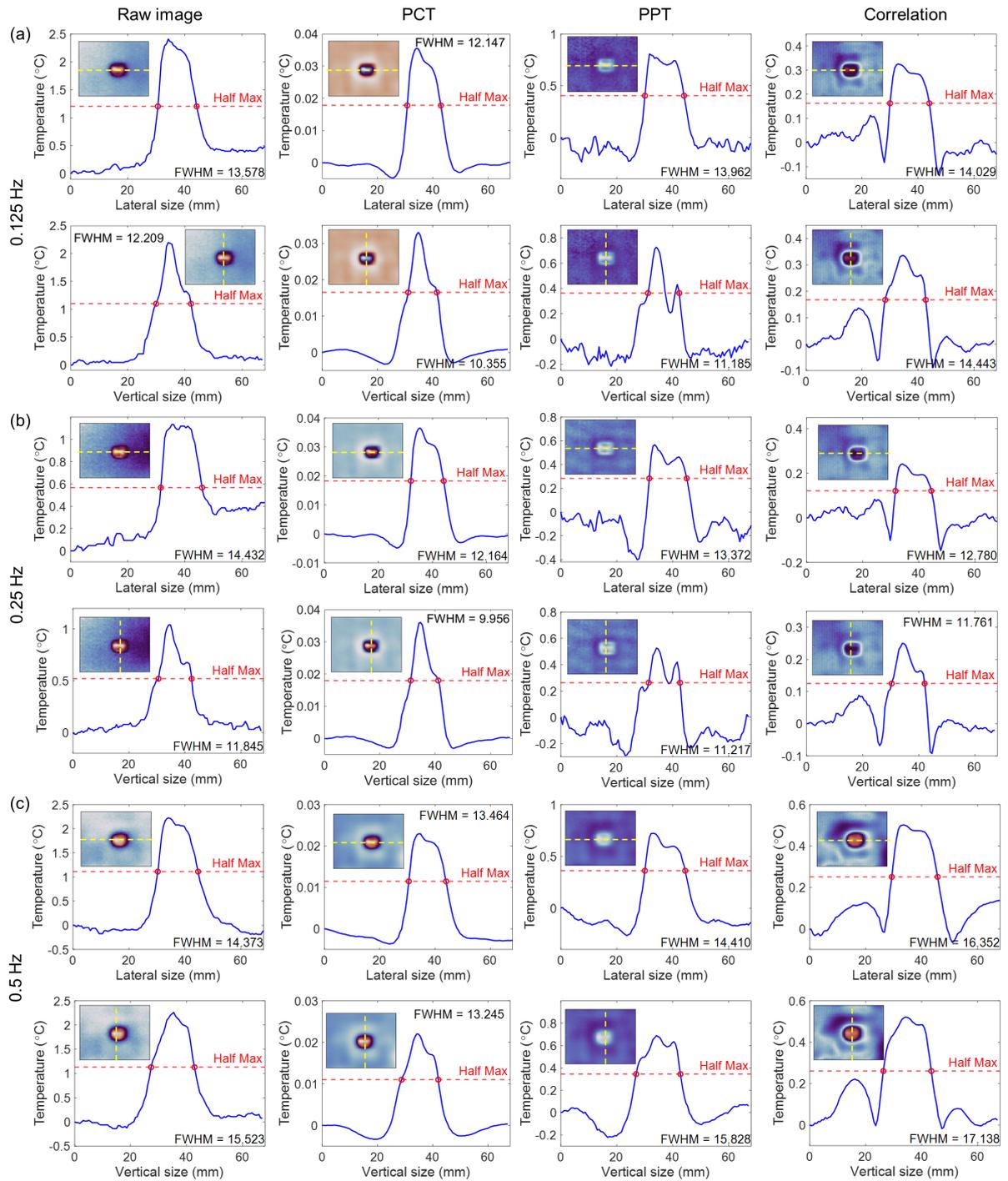

**Fig. 12.** Quantitative evaluation of the defect size for different image processing techniques under different envelope frequencies: (a) 0.125 Hz, (b) 0.25 Hz, (c) 0.5 Hz.

The quantitative evaluation results of the defect size based on virtual wave method are shown in Fig. 13. Comparing with PPT and correlation methods, virtual wave method has more stable baseline. The final quantitative results are shown in Table 3. It is clear to find that the proposed virtual wave method exhibits highest accuracy while both PCT and PPT techniques have lowest accuracy. According to the mean value of absolute errors, we can conclude that they follow a rank of Virtual wave (0.81) > Raw image (1.51) > Correlation (1.75) > PPT (1.95) > PCT (3.11).

Combined with three indices, we can find that the proposed virtual wave method has the best performance for defect detection, especially in SNR improvement and defect size quantification.

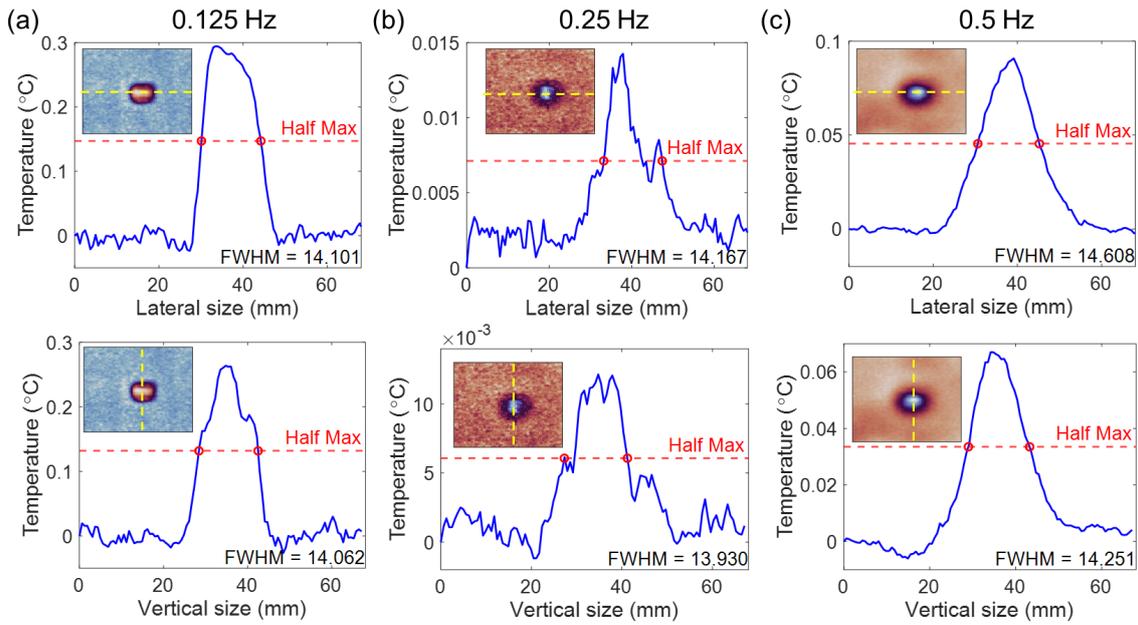

**Fig. 13.** Quantitative evaluation of the defect size for virtual wave method under different envelope frequencies: (a) 0.125 Hz, (b) 0.25 Hz, (c) 0.5 Hz.

**Table 3**
Quantitative evaluation of defect size for different image processing techniques. Unit: mm.

| Frequency | Direction | Raw image | PCT | PPT | Correlation | Virtual wave |
|---|---|---|---|---|---|---|
| 0.125 Hz | Lateral | 13.58 (-1.42) | 12.15 (-2.85) | 13.96 (-1.04) | 14.03 (-0.97) | **14.10 (-0.90)** |
|  | Vertical | 12.21 (-2.79) | 10.36 (-4.64) | 11.19 (-3.81) | **14.44 (-0.56)** | 14.06 (-0.94) |
| 0.25 Hz | Lateral | **14.43 (-0.57)** | 12.16 (-2.84) | 13.37 (-1.63) | 12.78 (-2.22) | 14.17 (-0.83) |
|  | Vertical | 11.85 (-3.15) | 9.96 (-5.04) | 11.22 (-3.78) | 11.76 (-3.24) | **13.93 (-1.07)** |
| 0.5 Hz | Lateral | 14.37 (-0.63) | 13.46 (-1.54) | 14.41 (-0.59) | 16.35 (+1.35) | **14.61 (-0.39)** |
|  | Vertical | **15.52 (+0.52)** | 13.25 (-1.75) | 15.83 (+0.83) | 17.14 (+2.14) | 14.25 (-0.75) |

*6.3. Photothermal tomography based on virtual wave reconstruction method*

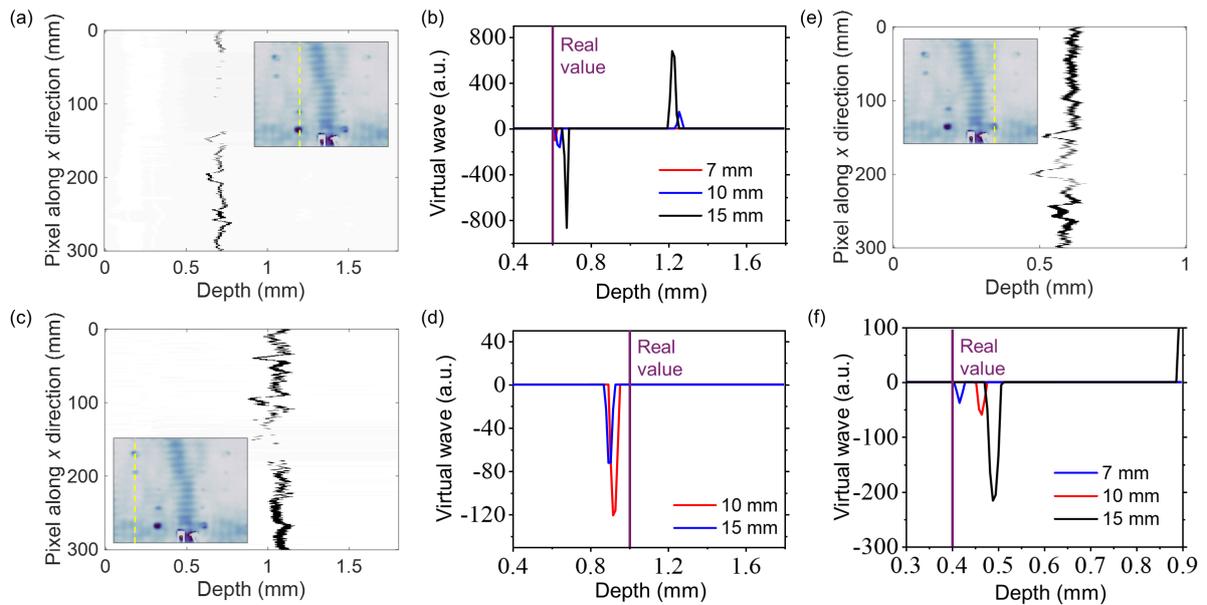

**Fig. 14.** Quantitative evaluation of the defect depth for virtual wave method under different 0.5 Hz excitation: (a) Depth tomogram at the second column, (b) Depth profile of different size defects at the second column, (c) Depth tomogram at the first column, (b) Depth profile of different size defects at the first column, (e) Depth tomogram at the fourth column, (f) Depth profile of different size defects at the fourth column.

Defect depth measurement or tomography is a hard issue in vibrothermography fields. This is because general lock-in vibrothermography has limited bandwidth comparing with optical excited thermography [51]. Furthermore, general infrared thermography techniques obtain subsurface information under the reflection mode, while vibrothermography depends on the defect itself therefore can be seen as an internal heat source. As our best knowledge, there is no open literature claiming that they could measure the subsurface defect depth or achieve photothermal tomography based on vibrothermography techniques. In this work, we solved this problem for the first time based on virtual wave method.

The virtual wave results are shown in Fig. 14. Different from results in the previous section, we processed the raw infrared data using the ADMM algorithm. The first, second, and fourth columns were selected to analysis and evaluate. According to results in Fig. 14(b), (d), and (f), it is clear to find the wavefront of reconstructed virtual wave signals. To quantitatively evaluate the accuracy of the proposed virtual wave method, we selected the depth values at different defect areas. The mean value of the depth information is shown in Table 4. With the increase of defect depth, the absolute error of virtual wave results decreases. This is because the thermal diffusion is related to the time. Therefore, the shallow defect appears at the early thermal sequence. Low frame rate of the infrared camera will lose these details of shallow defect depth although we can find these defects in the latter frames. According to the quantitative results in Table 4, the proposed virtual wave method demonstrates a high accuracy for defect depth detection.

**Table 4**
Quantitative evaluation of defect depth based on virtual wave method. Unit: mm.

| Location | 1st column | | 2nd column | | | 4th column | | |
|---|---|---|---|---|---|---|---|---|
| Defect size | 10 | 15 | 7 | 10 | 15 | 7 | 10 | 15 |
| Mean depth | 1.04 | 1.03 | 0.69 | 0.65 | 0.69 | 0.57 | 0.55 | 0.61 |
| Real depth | 1 | 1 | 0.6 | 0.6 | 0.6 | 0.4 | 0.4 | 0.4 |
| Absolute error | 0.04 | 0.03 | 0.09 | 0.05 | 0.09 | 0.17 | 0.15 | 0.21 |

## 7. Conclusions

In this work, a generalized virtual wave reconstruction framework for vibrothermography was proposed for addressing the inherent wavefront-free nature of thermal diffusion and the associated limitations in quantitative defect evaluation. Unlike conventional virtual wave formulations restricted to Dirac-type excitation, the proposed theory establishes a rigorous spatiotemporal relationship between thermal diffusion-wave fields and virtual wave field under arbitrary heat-generation conditions, without imposing simplifying assumptions on the underlying thermo-mechanical coupling. The resulting inversion problem is solved using two complementary regularization strategies – truncated singular-value decomposition (T-SVD) and the alternating direction method of multipliers (ADMM) – to ensure numerical stability and suppress noise amplification. Numerical simulations demonstrate that the reconstructed virtual wave fields recover wave-propagation characteristics that are absent in diffusive temperature distributions, thereby enhancing defect boundary definition improving contrast, and enabling depth-resolved analysis. Experiments conducted on CFRP laminates further confirm the robustness of the approach. When compared with established thermographic processing techniques (PCT, PPT, and correlation analysis), the proposed method achieves substantially higher signal-to-noise ratios, improved spatial clarity and the most accurate estimation of defect dimensions. Moreover, the framework enables, for the first time in vibrothermography, a tomographic assessment of defect depth through virtual wave inversion. The results indicate that the generalized virtual wave reconstruction method offers a consistent and effective pathway for quantitative vibrothermographic inspection.

## CRediT authorship contribution statement

**Pengfei Zhu:** Methodology, Investigation, Visualization, Writing – original draft, Writing – review & editing. **Julien Lecompagnon:** Methodology, Writing – review & editing, **Mathias Ziegler:** Project administration, Funding acquisition, **Clemente Ibarra-Castanedo:** Resources, Data curation, Writing – review & editing. **Xavier Maldague:** Project administration, Funding acquisition.

## Declaration of Competing Interest

The authors declare that they have no known competing financial interests or personal relationships that could have appeared to influence the work reported in this paper.

## Acknowledgments

This work was supported by the Adolf Marten Fellowship (Grant n. BAM-AMF-2025-1), and Natural Sciences and Engineering Research Council (NSERC) of Canada through the Discovery and CREATE 'oN DuTy!' program (496439-2017), the Canada Research Chair in Multipolar Infrared Vision (MiViM).

**Data availability**

Data will be made available on request.